\documentclass[fleqn,usenatbib]{mnras}

\usepackage[T1]{fontenc}
\usepackage{ae,aecompl}

\usepackage[dvipdfmx]{graphicx}	
\usepackage{amsmath}	
\usepackage{amssymb}	
\usepackage[hang,small,bf]{caption}
\usepackage[subrefformat=parens]{subcaption}
\usepackage{color}
\usepackage{bm}

\title[Effect of radiation pressure]{The effect of radiation pressure on spatial distribution of dust inside H$_\mathrm{II}$ regions}

\author[S. Ishiki, T. Okamoto \& A. K. Inoue]{
Shohei Ishiki,$^{1}$\thanks{E-mail: ishiki@astro1.sci.hokudai.ac.jp}
Takashi Okamoto,$^{1}$
and Akio K. Inoue$^{2}$
\\
$^{1}$Department of Cosmoscience, Hokkaido University, N10 W8, Kitaku, Sapporo, 060-0810, Japan\\
$^{2}$Department of Environmental Science and Technology, Faculty of Design Technology, \\
Osaka Sangyo University, 3-1-1 Nakagaito, Daito, Osaka 574-8530, Japan\\
}

\date{Accepted XXX. Received YYY; in original form ZZZ}

\pubyear{2015}

\begin{document}
\label{firsPHage}
\pagerange{\pageref{firsPHage}--\pageref{lasPHage}}
\maketitle

\begin{abstract}
We investigate the impact of radiation pressure on spatial dust distribution inside \ion{H}{ii} regions 
using one-dimensional radiation hydrodynamic simulations, which include 
absorption and re-emission of photons by dust. 
In order to investigate grain size effects as well, we introduce two additional fluid components 
describing large and small dust grains in the simulations. 
Relative velocity between dust and gas strongly depends on the drag force. We include 
collisional drag force and coulomb drag force. 
  We find that, in a compact \ion{H}{ii} region, a dust cavity region is formed by radiation pressure. 
  Resulting dust cavity sizes ($\sim 0.2$~pc) agree with observational estimates reasonably well. 
  Since dust inside an \ion{H}{ii} region is strongly charged, relative velocity between dust and gas is mainly determined by the coulomb drag force. 
Strength of the coulomb drag force is about 2-order of magnitude larger than that of the collisional drag force. 
  In addition, in a cloud of mass $10^5$~$M_{\sun}$, we find that the radiation pressure 
  changes the grain size distribution inside \ion{H}{ii} regions. 
  Since large (0.1 \micron) dust grains are accelerated more efficiently than small (0.01 \micron) grains, 
  the large to small grain mass ratio becomes smaller by an order of magnitude  
  compared with the initial one. Resulting dust size distributions depend on the luminosity of the radiation source. 
  The large and small grain segregation becomes weaker when we assume 
  stronger radiation source, since dust grain charges become larger under stronger radiation and hence coulomb drag force
   becomes stronger. 
  \end{abstract}




\section{Introduction}

Radiation from young massive stars plays a crucial role in star forming regions, and its effect on spatial dust distribution inside \ion{H}{ii} regions is also non-negligible. 
\cite{ODell1965} firstly observed dust inside the \ion{H}{ii} region and  many other observations found dust in \ion{H}{ii} regions \citep{ODell1966, Kawajiri1968, Harper1971}.
\cite{ODell1965} observationally estimated the distribution of dust inside \ion{H}{ii} regions, concluding that gas-to-dust mass ratio decreases as 
a function of distance from the centre of the nebulae. \cite{Nakano1983} and \cite{Chini1987} observationally suggested the existence of  dust cavity regions. There have been some theoretical attempts to reveal dust distribution inside \ion{H}{ii} regions \citep{Mathews1967, Gail1979a, Gail1979b}. \cite{Gail1979b} suggested that a dust cavity can be created by radiation pressure. 

Radiation pressure may also produce spatial variations in the grain size distribution inside \ion{H}{ii} regions 
as suggested by recent observational data of IR bubbles. 
From the Galactic Legacy Ingrared Mid-Plane Survey Extraordinaire \citep[GLIMPSE;][]{Benjamin2003}, 
\cite{Churchwell2006} found that about 25\% of IR bubbles are associated with known \ion{H}{ii} regions and they claimed that the IR bubbles are primarily formed around hot young stars. 
\cite{Deharveng2010} then pointed out that 86\% of IR bubbles are associated with ionzed gas. 
Since \cite{Churchwell2006} missed the large (> 10 arcmin) and small (< 2 arcmin) bubbles, 
\cite{Simpson2012} presented a new catalogure of 5106 IR bubbles. 
\cite{Paladini2012} found that the peak of 250~\micron\, continuum emission appears further from radiation source than that of 8~\micron\, continuum emission. 
Since they assumed that 250~\micron\, continuum emission traces the big grains (BGs) and 8~\micron\, continuum emission traces the polycyclic aromatic hydrocarbons (PAHs), 
they argued that the dust size distribution depends on the distance from a radiation source. 
 
%
\cite{Inoue2002} argued the presence of the central dust depleted region --- dust cavity --- in compact/ultra-compact \ion{H}{ii} regions in the Galaxy by comparing the observed infrared-to-radio flux ratios with a simple spherical radiation transfer model. The dust cavity radius is estimated to be 30\% of the Stromgren radius on average, which is too large to be explained by dust sublimation. The formation mechanism of the cavity is still an open question, while the radiation pressure and/or the stellar wind from the excitation stars have been suggested as responsible mechanisms. We will examine whether the radiation pressure can produce the cavity in this paper.
 By considering the effect of radiation pressure on dust and assuming steady \ion{H}{ii} regions, 
 \cite{Draine2011} theoretically explained the dust cavity size that \cite{Inoue2002} estimated from observational data. 
 
 \cite{Akimkin2015,Akimkin2017} estimated dust size distribution by solving motion of dust and gas respectively, 
 and they concluded that radiation pressure preferentially removes large dust from \ion{H}{ii} regions.   
 Their simulations have, however, assumed a single OB star as a radiation source. As mentioned by \cite{Akimkin2015}, 
 grain electric potential is the main factor that affects the dust size distribution.  
 If we assume a stronger radiation source, such as a star cluster, dust would been more strongly charged and their conclusions might change.
 

In this paper, 
we investigate the effect of radiation pressure on spatial dust distribution inside compact \ion{H}{ii} regions 
and compare it with the observational estimates \citep{Inoue2002}. 
In addition, 
we perform multi-dust-size simulations and study the effect of the luminosity of the radiation source on dust size distribution inside \ion{H}{ii} regions. 

The structure of this paper is as follows:
In Section~\ref{sec:Methods}, we describe our simulations. 
In Section~\ref{sec:setup}, we describe our simulation setup.
In Section~\ref{sec:Results}, we present simulation results. 
In Section~\ref{sec:Discussion}, we discuss the results and present our conclusions.

\section{Methods}
\label{sec:Methods}

We place a radiation source at the centre of a spherically symmetric gas distribution. 
The species we include in our simulations are \ion{H}{i}, \ion{H}{ii}, \ion{He}{i}, \ion{He}{ii}, 
\ion{He}{iii}, electrons, and dust. 
We assume the dust-to-gas mass ratio to be $6.7 \times 10^{-3}$ corresponding to a half of 
the abundance of elements heavier than He (so-called 'metal') in the Sun \citep{asplund09}. 
We neglect gas-phase metal elements in this paper.
We solve the radiation hydrodynamic equations at each timestep as follows:
\begin{enumerate}
 \item[step 1] Hydrodynamic equations
 \item[step 2] Radiative transfer and other related processes
 \begin{enumerate}
 \item[substep 2.1] Static radiative transfer equations
 \item[substep 2.2] Chemical reactions
 \item[substep 2.3] Radiative heating and cooling
 \item[substep 2.4] Grain electric potential
 \end{enumerate}\end{enumerate} 
\noindent The methods we use for radiation transfer, chemical reactions, radiative heating, cooling and time stepping 
are the same as \citet[hereafter paper I]{Ishiki2017}.

\subsection{Dust model}

We include absorption and thermal emission of photons by dust grains in our simulations. 
To convert the dust mass density to the grain number density, we assume a graphite grain whose material density is 
2.26\,g\,cm$^{-3}$ \citep{Draine1979}. 
We employ the cross-sections of  dust in \citet{1984ApJ...285...89D} 
and \citet{1993ApJ...402..441L}\footnote{http://www.astro.princeton.edu/\textasciitilde draine/dust/dust.diel.html}. 
Dust sizes we assume are 0.1 \micron\, or 0.01 \micron. 
Dust temperature is determined by the radiative equilibrium, and thus  
the dust temperature is independent from gas temperature. 
We assume that the dust sublimation temperature is $1500$~K; however, 
dust never be heated to this temperature in our simulations. 
We do not include photon scattering by dust grains for simplicity.

\subsection{Grain electric potential}

In our simulations, we solve hydrodynamics including the coulomb drag force 
which depends on grain electric potential. 
In order to determine the grain electric potential, we consider following processes: 
primary photoelectric emission, auger electron emission, secondary electron emission, and electron and ion collisions \citep{Weingartner2001, Weingartner2006}. 
The effect of auger electron emission and secondary electron emission is, however, 
almost negligible in our simulations, 
because high energy photons ($> 10^2$~eV) responsible for the two processes are 
negligible in the radiation sources considered in this paper.
Since the time scale of dust charging processes is so small ($\lesssim$1 yr), 
we integrate the equation of grain electric potential implicitly. 

\subsection{Dust drag force}

In our simulations, we calculate the effect of drag force $F_\mathrm{drag}$ on a dust of charge $Z_\mathrm{d}$ and radius $a_\mathrm{d}$ \citep{Draine1979} as follows: 
 \begin{equation}
 F_\mathrm{drag} = 2 \pi a_\mathrm{d}^2 k T_\mathrm{g} \left[ \sum_\mathrm{i} n_\mathrm{i} \left( G_0 (s_\mathrm{i}) + z_\mathrm{i}^2 \phi^2 \mathrm{ln} (\Lambda / z_\mathrm{i}) G_2 (s_\mathrm{i}) \right) \right] , \nonumber 
 \end{equation}
 where
 \begin{eqnarray}
 s_\mathrm{i} \equiv \sqrt{m_\mathrm{i} v^2 / (2kT_\mathrm{g})}, \nonumber \\
 G_0 (s_\mathrm{i}) \approx 8s_\mathrm{i} / (3 \sqrt{\pi} ) \sqrt{1+9\pi s_\mathrm{i}^2/64}, \nonumber \\
 G_2 (s_\mathrm{i}) \approx s_\mathrm{i} / (3 \sqrt{\pi}/4 + s_\mathrm{i}^3), \nonumber \\
 \phi \equiv Z_\mathrm{d} e^2 / ( a_\mathrm{d} k T_\mathrm{g} ), \nonumber \\
 \Lambda \equiv 3/ (2 a_\mathrm{d} e | \phi |) \sqrt{k T_\mathrm{g}/ \pi n_\mathrm{e}}, \nonumber
 \end{eqnarray}
 $k$ is the Boltzmann constant, $T_\mathrm{g}$ is the temperature of gas, $n_\mathrm{i}$ is the number density of $i$th gas species, $n_\mathrm{e}$ 
 is the number density of electron, $z_\mathrm{i}$ is the charge of $i$th gas species ($i=$ \ion{H}{i}, \ion{H}{ii}, \ion{He}{i}, \ion{He}{ii}, \ion{He}{iii}), and $m_\mathrm{i}$ is the mass of $i$th species. 

\subsection{Hydrodynamics}

\subsubsection{dust and gas dynamics}
\label{subsubsec:d1} 

In this section we describe the procedure to solve the set of hydrodynamic equations:
\begin{eqnarray}
\frac{\partial}{\partial t} \rho_\mathrm{g} + \frac{\partial}{\partial x} \rho_\mathrm{g} v_\mathrm{g} &=& 0 \nonumber \\ 
\frac{\partial}{\partial t} \rho_\mathrm{d} + \frac{\partial}{\partial x} \rho_\mathrm{d} v_\mathrm{d} &=& 0 \nonumber \\ 
\frac{\partial}{\partial t} \rho_\mathrm{g} v_\mathrm{g} + \frac{\partial}{\partial x} \rho_\mathrm{g} v_\mathrm{g}^2 &=& \rho_\mathrm{g} a_\mathrm{gra} + f_\mathrm{rad,g} - \frac{\partial}{\partial x} P_\mathrm{g} \nonumber \\ 
& & + K_\mathrm{d} (v_\mathrm{d} - v_\mathrm{g}) \nonumber \\
\frac{\partial}{\partial t} \rho_\mathrm{d} v_\mathrm{d} + \frac{\partial}{\partial x} \rho_\mathrm{d} v_\mathrm{d}^2 &=&  \rho_\mathrm{d} a_\mathrm{gra} + f_\mathrm{rad,d}  \nonumber \\ 
& & + K_\mathrm{d} (v_\mathrm{g} - v_\mathrm{d}) \nonumber \\
\frac{\partial}{\partial t} \left( \frac{1}{2} \rho_\mathrm{g} v_\mathrm{g}^2 +  \frac{1}{2} \rho_\mathrm{d} v_\mathrm{d}^2 + e_\mathrm{g} \right) &+& \frac{\partial}{\partial x} \left[ \left( \frac{1}{2} \rho_\mathrm{g} v_\mathrm{g}^2 + h_\mathrm{g} \right) v_\mathrm{g} + \frac{1}{2} \rho_\mathrm{d} v_\mathrm{d}^3 \right] \nonumber \\
&=& \left( \rho_\mathrm{g} v_\mathrm{g} +  \rho_\mathrm{d} v_\mathrm{d} \right) a_\mathrm{gra} \nonumber \\
& & + f_\mathrm{rad,g} v_\mathrm{g} + f_\mathrm{rad,d} v_\mathrm{d} \nonumber 
\end{eqnarray}
where $\rho_\mathrm{g}$ is the mass density of gas, $\rho_\mathrm{d}$ is the mass density of dust, $v_\mathrm{g}$ is the velocity of gas, $v_\mathrm{d}$ is 
the velocity of dust, $a_\mathrm{gra}$ is the gravitational acceleration, $f_\mathrm{rad,g}$ is the radiation pressure 
gradient force on gas, $f_\mathrm{rad,d}$ is the radiation pressure gradient force on dust, 
$P_\mathrm{g}$ is the gas pressure, $e_\mathrm{g}$ is the internal energy of gas, $h_\mathrm{g}$ is the enthalpy of gas, and $K_\mathrm{d}$ is the drag coefficient between gas and dust defined as follows:  
 \begin{equation}
 K_\mathrm{d} \equiv \frac{n_\mathrm{d} F_\mathrm{drag}}{|\bm{v}_\mathrm{d} - \bm{v}_\mathrm{g}|}, \nonumber
 \end{equation}
 where $n_\mathrm{d}$ is the number density of dust grains. 

In order to solve the dust drag force stably, we use following algorithm for the momentum equations:
\begin{equation}
\begin{split}
\begin{bmatrix}
p_\mathrm{d}^{*} \left( = \rho_\mathrm{d}^\mathrm{t+\Delta t} v_\mathrm{d}^*  \right) \\
p_g^{*}\left( = \rho_\mathrm{g}^\mathrm{t+\Delta t} v_\mathrm{g}^*  \right) 
\end{bmatrix}
&=
\begin{bmatrix}
p_\mathrm{d}^t \\
p_g^t 
\end{bmatrix}
+ \begin{bmatrix}
F_\mathrm{p, d}(\rho_\mathrm{d}^t,v_\mathrm{d}^t)  \\
F_\mathrm{p, g}(\rho_g^t,v_\mathrm{g}^t,e_g^t) 
\end{bmatrix}
\Delta t,\label{cdv1}
\end{split}
\end{equation}
\begin{equation}
\begin{split}
\begin{bmatrix}
p_\mathrm{d}^\mathrm{t+\Delta t} \\
p_\mathrm{g}^\mathrm{t+\Delta t}
\end{bmatrix}
&=
\begin{bmatrix}
\rho_\mathrm{d}^\mathrm{t+\Delta t} \\
\rho_\mathrm{g}^\mathrm{t+\Delta t} 
\end{bmatrix}
\frac{ p_\mathrm{d}^*+p_\mathrm{g}^* }{ \rho_\mathrm{d}^\mathrm{t+\Delta t} + \rho_g^\mathrm{t+\Delta t}  } \\
&+
\begin{bmatrix}
\rho_\mathrm{d}^\mathrm{t+\Delta t} \\
\rho_\mathrm{g}^\mathrm{t+\Delta t} 
\end{bmatrix}
\left[ a_\mathrm{gra} + \frac{f_\mathrm{d} + f_g }{\rho_\mathrm{d}^\mathrm{t+\Delta t} + \rho_g^\mathrm{t+\Delta t} }\right] \Delta t  \label{cdv2} \\
&+ 
\begin{bmatrix}
-1 \\
1
\end{bmatrix}
\frac{\rho_g^\mathrm{t+\Delta t} \rho_d^\mathrm{t+\Delta t}}{\rho_g^\mathrm{t+\Delta t} + \rho_d^\mathrm{t+\Delta t}}  \left( v_\mathrm{g}^* - v_\mathrm{d}^* \right) \mathrm{e}^{-\frac{\Delta t}{t_d}} \\
&+ 
\begin{bmatrix}
-1 \\
1
\end{bmatrix}
t_d  \frac{\rho_d^\mathrm{t+\Delta t} f_g - \rho_g^\mathrm{t+\Delta t} f_d}{\rho_g + \rho_d} ( 1 - \mathrm{e}^{-\frac{\Delta t}{t_d}} ),  
\end{split} 
\end{equation}
where $\Delta t$ is the time step, $\rho_\mathrm{i}^\mathrm{t}$ is the mass density of $i$th species at time $t$, $p_\mathrm{i}^\mathrm{t}$ is the momentum of $i$th species at time $t$, $e_\mathrm{g}^\mathrm{t}$ is the internal energy of gas at time $t$, $F_\mathrm{X,i}$ is the advection of the physical quantity $X$ of the $i$th species, $f_\mathrm{d}$ is  the force on dust ($f_\mathrm{d}=f_\mathrm{rad,d}$), $f_\mathrm{g}$ is the force on gas ($f_\mathrm{g}=f_\mathrm{rad,g}-\partial P_\mathrm{g}/ \partial x$), and the inverse of the drag stopping time, $t_\mathrm{d}$, is
\begin{equation} 
t_\mathrm{d}^{-1} =  K_\mathrm{d} \frac{\rho_\mathrm{d}^\mathrm{t+\Delta t} + \rho_\mathrm{g}^\mathrm{t+\Delta t} }{\rho_\mathrm{d}^\mathrm{t+\Delta t}  \rho_\mathrm{g}^\mathrm{t+\Delta t} }. \label{td}
\end{equation}
Equation (\ref{cdv2}) that determines the relative velocity between dust and gas is the 
exact solution of the following equations:
\begin{equation}
\begin{split}
\rho_\mathrm{g} \frac{d}{dt} v_\mathrm{g} &= f_\mathrm{rad,g} - \frac{\partial}{\partial x}P_\mathrm{g} + \rho_\mathrm{g} a_\mathrm{gra} + K_\mathrm{d} (v_\mathrm{d} - v_\mathrm{g}), \label{dve} \\
\rho_\mathrm{d} \frac{d}{dt} v_\mathrm{d} &= f_\mathrm{rad,d}  + \rho_\mathrm{d} a_\mathrm{gra} + K_\mathrm{d} (v_\mathrm{g} - v_\mathrm{d}). 
\end{split}
\end{equation}

Momentum advection and other hydrodynamic equations are solved by using {\small \textsc{AUSM$+$}} \citep{1996JCoPh.129..364L}. 
We solve the hydrodynamics in the second order accuracy in space and time. 
In order to prevent cell density from becoming zero or a negative value, 
we set the minimum number density, $n_\mathrm{H} \simeq 10^{-13}$~cm$^{-3}$. 
We have confirmed that our results are not sensitive to the choice of the threshold 
density as long as the threshold density is sufficiently low. 

In order to investigate whether our method is reliable, we perform shock tube tests in Appendix \ref{sec:stt}. 
ln Appendix \ref{sec:2dust}, we describe how we deal with the dust grains with two sizes.


\section{Simulation setup}
\label{sec:setup} 
\begin{table*}
  \centering
  \caption{
  Initial conditions and numerical setup for the simulations. The radius of 
    each cloud are shown as $r_\mathrm{cloud}$. 
    The number densities, $n_\mathrm{H}$, $n_\mathrm{He}$, and $n_\mathrm{d}$ indicate the initial
    number densities of hydrogen, helium, and dust in the innermost cell, respectively.  
    Spatial distribution of dust and gas are indicated by `C' (constant density) and `BE' (Bonnor-Ebert sphere) 
    in the 7th column. 
    The spectrum of a radiation source is shown in the 8th column, where `SSP' and `BB' respectively indicate 
    a simple stellar population with solar metallicity and the black body spectrum of given temperature. 
    $\dot{N}_\mathrm{ion}$ represents the number of ionized photon emitted from a radiation source per unit time.
    The initial temperature of gas and dust are represented by $T_\mathrm{g}$ and $T_\mathrm{d}$, 
    respectively. 
    The mass of a central radiation source is indicated by 
    $M_\mathrm{star}$. 
    The number of dust sizes is shown in the 13th column. 
   When `Gravity' is on/off, we include/ignore gravitation force in the simulations. 
  }
  \scalebox{0.85}[0.85]{
\begin{tabular}{lccccccccccccc} \hline
  Cloud & $r_\mathrm{cloud}$ & ${n}_\mathrm{{\mathrm{H}}}$ & ${n}_\mathrm{{\mathrm{He}}}$ & ${n}_\mathrm{{\mathrm{d,Large}}}$ & ${n}_\mathrm{{\mathrm{d,Small}}}$ & Distribution & Source & $\dot{N}_\mathrm{ion}$ & ${T_\mathrm{\mathrm{g}}}$ & ${T_\mathrm{d}}$ & ${M}_\mathrm{{\mathrm{star}}}$ & Dust & Gravity \\ 
   & (pc) & (cm$^{-3}$) & (cm$^{-3}$) & ($10^{-10}$~cm$^{-3}$) & ($10^{-7}$~cm$^{-3}$) & & & ($10^{49}$~s$^{-1}$) & (K) & (K) & ($10^3~M_\mathrm{{\sun}}$) &  \\ \hline
  Cloud 1 &1.2 & 4$\times$10$^5$ & 3.4$\times$10$^4$ & 6.4$\times$10$^{3}$ & 0 & C & BB (50100K) &  6.2 & 100 & 10 & 0.08 & 1 & off \\ 
  Cloud 2 &17 & 791 & 67 & 9.6 & 3.0 & BE & BB (38500K) & 0.72 & 1082 & 10 & 0.05 & 2 & on \\ 
  Cloud 3 &17 & 791 & 67 & 9.6 & 3.0 & BE & SSP & 5.8 & 1082 & 10 & 2 & 2 & on \\ 
  Cloud 4 &17 & 791 & 67 & 9.6 & 3.0 & BE & SSP & 58 & 1082 & 10 & 20 & 2 & on \\  \hline
   \end{tabular}
   }
  \label{Tab1}
\end{table*}
  In the first simulation, in order to investigate whether our simulation derives a consistent result with the observational estimate for compact/ultra-compact \ion{H}{ii} regions \citep{Inoue2002},  
we model a constant density cloud of hydrogen number density $4\times10^5$~cm$^{-3}$ and radius 1.2~pc. 
As a radiation source, we  place a single star (i.e. black body) at the centre of the sphere. 
Since we are interested in the formation of a dust cavity, we neglect the gravity 
which does not affect the relative velocity between dust grains and gas 
(see equation~(\ref{cdv2})). 
We assume a single dust grain size in this simulation. 

In the second set of simulations, in order to investigate the effect of radiation pressure on the dust grain size distribution inside a large gas cloud, 
we model a cloud as a Bonner-Ebert sphere of mass $10^5$~$M_{\sun}$ and radius 17~pc. 
As the radiation source, we consider a single star (black body, BB) or a star cluster (a simple stellar population, SSP) 
and we change the luminosity of the radiation source
to investigate the dependence of the dust size distribution on the luminosity of the radiation source. 
We compute its luminosity and spectral-energy distribution as a function of time by using a population synthesis code, {\small \textsc{P\'{E}GASE.2}} \citep{1997A&A...326..950F, 1999astro.ph.12179F}, assuming the Salpeter initial mass function \citep{salpeter} and the solar metallicity. We set the mass range of the initial mass function to be 0.1 to 120~$M_{\sun}$.

Materials at radius, $r$, feel the radial gravitational acceleration, 
\begin{equation}
a_\mathrm{gra}(r) = - G \frac{M_\mathrm{star} r}{(r^2 + r_\mathrm{soft}^2)^{3/2}}- G \frac{M (<r) }{r^2 } \nonumber
\end{equation}
where M(<r) represents the total mass of gas inside $r$ and $M_\mathrm{star}$ is the mass of the central radiation source,  
which is 50~$M_{\sun}$ for the single star case and $2\times10^3$ or $2\times10^4$~$M_{\sun}$ for the two star cluster cases.
Since the gravity from the radiation source has a non-negligible effect on simulation results and causes numerical instability in the case of SSP,  we need to introduce softening length, $r_{\mathrm{soft}}$.
We set it to 0.5~pc for the SSP. Since the gravity from a single star is negligible effect on simulation results, we set 0~pc for the single star. 

Following the dust size distribution of \cite{MRN}, so-called MRN distribution, we 
assume two dust size in these simulations. We assume the initial number ratio of large to small 
dust as 
\begin{equation}
n_{\mathrm{d, Large}} : n_\mathrm{d, Small} = 1 : 10^{2.5}, \nonumber
\end{equation}
where $n_{\mathrm{d, Large}}$ and $n_\mathrm{d, Small}$ are the number density of dust grains of 0.1~\micron\,  and 0.01~\micron\, in size, 
respectively.

The details of initial conditions are listed in Table~\ref{Tab1}.
We use linearly spaced 128 meshes in radial direction, 128 meshes in angular direction, 
and 256 meshes in frequency direction in all simulations to solve radiation hydrodynamics. 

\section{Results}
\label{sec:Results}
\subsection{Dust cavity radius}
\begin{figure*}
\centering 
\includegraphics[width=8.0cm]{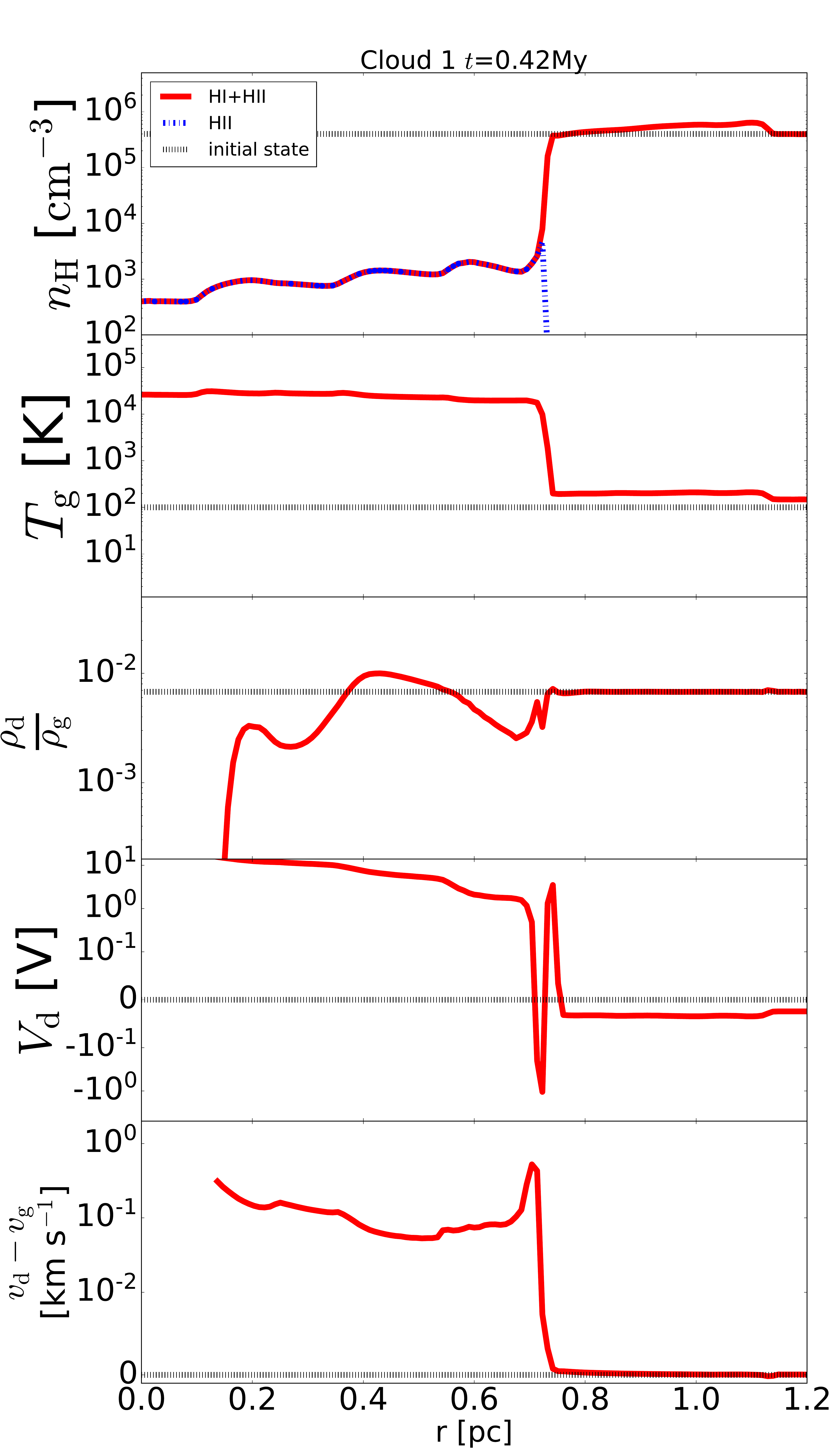} 
\caption{
  Density (top-row), gas temperature (second-row), dust-to-gas mass ratio (third-row), grain electric potential (fourth-row), 
  and relative velocity between dust and gas (bottom-row) profile at $t = 0.42$~Myr.  
  The black dotted lines show the initial profiles. 
  The red solid lines represent the simulation results.  
  The blue dashed lines in the top panel shows the ionized hydrogen density profile.   
} 
\label{d1g} 
\end{figure*}

\begin{table*}
  \centering
  \caption{
  Comparison between the simulation results ($t=0.42$~Myr) and the observational estimates. 
    The number densities of ionized electrons inside \ion{H}{ii} regions are represented by $\overline{n}_\mathrm{e}$. 
    The number of ionized photons emitted from radiation sources per unit time is represented by $\dot{N}_\mathrm{ion}$.
    The radius of \ion{H}{ii} region is represented by $r_\mathrm{\ion{H}{ii}}$. 
    The radius of dust cavity radius is represented by $r_\mathrm{d}$. 
    The parameter $y_\mathrm{d}$ is defined by $y_\mathrm{d} \equiv r_\mathrm{d}/R_\mathrm{St}$, where  $R_\mathrm{St}$ is the Str\"{o}mgren radius. 
    Since, observationally, the number density is driven from the column density, the electron number density in the simulation is defined as 
    by $\overline{n}_\mathrm{e} \equiv \int_0^{r_\mathrm{\ion{H}{ii}}} n_\mathrm{e} dx/r_\mathrm{\ion{H}{ii}}$. 
    In addition, in order to match the definition of $r_\mathrm{d}$ with \protect\cite{Inoue2002}, the dust cavity radius is 
    defined by $r_\mathrm{d} \equiv  r_\mathrm{\ion{H}{ii}} \left(1 - \int_0^{r_\mathrm{\ion{H}{ii}}} \rho_\mathrm{d}  dx/\int_0^{r_\mathrm{\ion{H}{ii}}} \rho_\mathrm{g} (\rho_\mathrm{d}/\rho_\mathrm{g})_\mathrm{initial} dx\right)$, where $(\rho_\mathrm{d}/\rho_\mathrm{g})_\mathrm{initial}$ represents the initial dust-to-gas mass ratio. 
  }
\begin{tabular}{lccccc} \hline
  \, & $\overline{n}_\mathrm{e}$ & $\dot{N}_\mathrm{ion}$ & ${r}_\mathrm{\ion{H}{ii}}$ & ${r}_\mathrm{d}$ & ${y}_\mathrm{d}$\\ 
  \, & (cm$^{-3}$) & (10$^{49}$s$^{-1}$) & (pc) & (pc) & $\equiv r_\mathrm{d}/R_\mathrm{St}$ \\ \hline
  this work ($t=0.42$~Myr) & 1247 & 6.2 & 0.73 & 0.21 & 0.20 \\ 
  \cite{Inoue2002} &1200 $\pm$ 400 & 6.8 $\pm$ 3.9 & 0.72 $\pm$ 0.098 & 0.28 $\pm$ 0.13 & 0.30 $\pm$ 0.12 \\ \hline
   \end{tabular}
  \label{Tab2}
\end{table*}

We present density, gas temperature, dust-to-gas mass ratio, grain electric potential ($V_\mathrm{d} \equiv e Z_d/a_\mathrm{d}$), and relative velocity between dust and gas as functions of radius in Fig \ref{d1g}. 
 In the top panels, the hydrogen number density is indicated by the red solid line. The number density of \ion{H}{ii} is indicated 
 by the blue dash-dotted line. The initial state of the simulation is shown by the black dotted line. 


The average electron number density within an \ion{H}{ii} region, $\overline{n}_\mathrm{e}$, the \ion{H}{ii} region radius, $r_\mathrm{\ion{H}{ii}}$, the dust cavity radius, $r_\mathrm{d}$, and the ratio between the radius of the dust cavity to the Str\"{o}mgren radius, $y_\mathrm{d}$, obtained by our simulation ($t=0.42$~Myr) and the observational estimate are shown in Table \ref{Tab2}. 
We find that our simulation results are in broad agreement with the observational estimate. 
The dust cavities, hence, could be created by radiation pressure. 
The parameter $y_\mathrm{d}$ obtained by the simulation is somewhat smaller than the observational estimate. However, we could find a better agreement if we tuned the initial condition such as the gas density. In addition, the agreement would be better if we included the effect of stellar winds, which was neglected in this paper.

Since dust inside the \ion{H}{ii} region is strongly charged, relative velocity between dust and gas is determined by coulomb drag force. 
Magnitude of the coulomb drag force is about 2-order of magnitude larger than that of the collisional drag force. 
The relative velocity, thus, becomes largest when the dust charge is neutral. 

Grain electric potential gradually decreases with radius and then suddenly drops to negative value. Near the ionization front, the number of ionized photons decreases and 
hence collisional charging becomes important. This is the reason behind the sudden decrease of the grain electric potential.
In the neutral region, there is no photon which is able to ionize the gas and hence there is few electrons that collide with dust grains. On the other hand, there are photons that 
photoelectrically charge dust grains. Therefore, the grain electric potential becomes positive again at just 
outside of the \ion{H}{ii} region. 
Then, the UV photons are consumed and the electron collisional charging becomes dominant again 
in the neutral region.

\subsection{Spatial distribution of large dust grains and small dust grains}

\begin{figure*}
\centering 
\includegraphics[width=16.5cm]{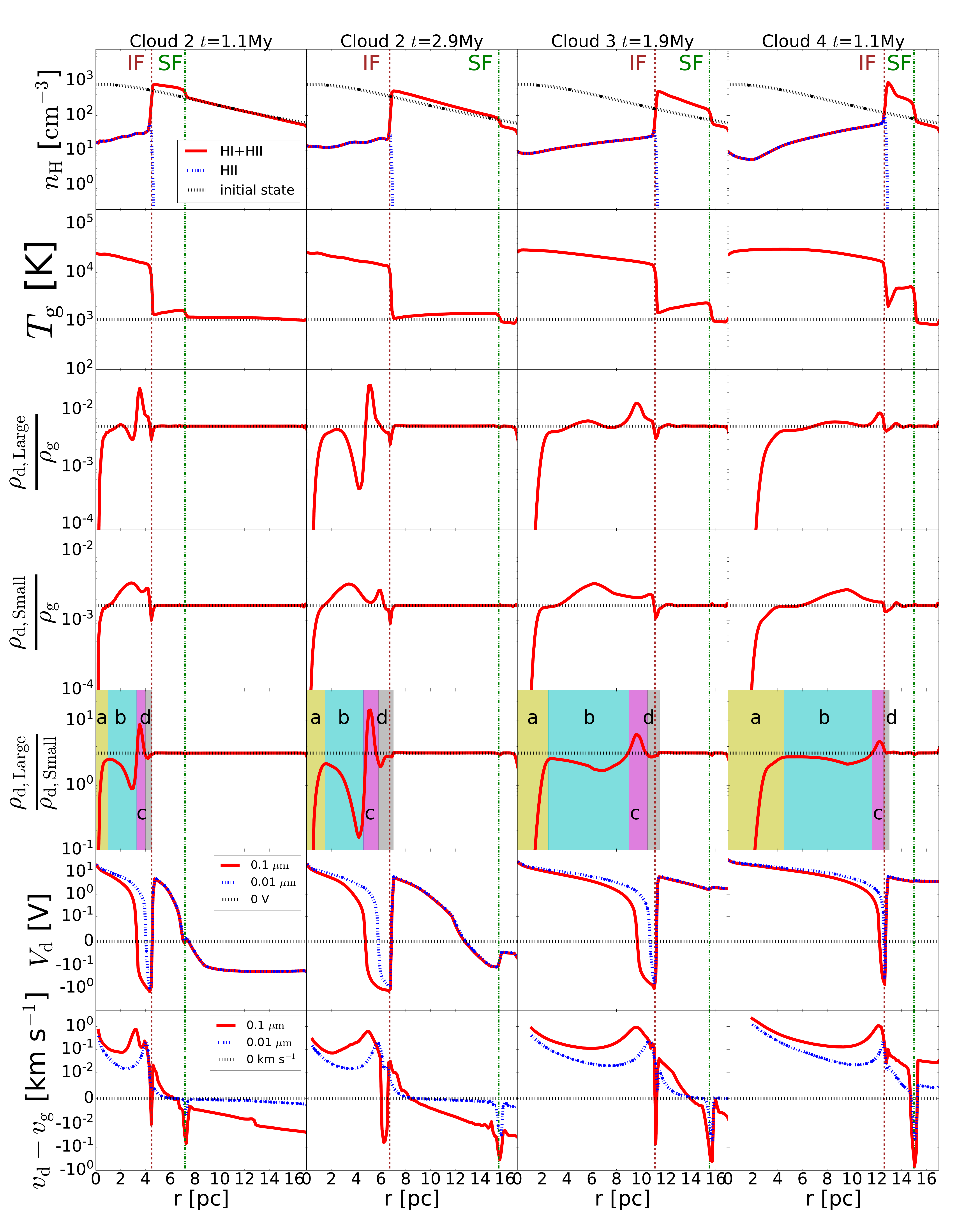} 
\caption{
  Density (top-row), gas temperature (second-row), large-dust-to-gas-mass ratio (third-row), small-dust-to-gas-mass ratio (fourth-row), 
  large-dust-to-small-dust-mass ratio (fifth-row), grain electric potential (sixth-row), and relative velocity between dust and gas (bottom-row) profiles.  
  From left to right, we show the results for Clouds~2 at $t = 1.1$~Myr, Cloud~2 at $t = 2.9$~Myr, Cloud~3 at $t = 1.9$~Myr, and Cloud~4 at $t = 1.1$~Myr. 
  The black dotted lines show the initial profiles. 
  The red solid lines represent the results of simulations.  
  The blue dashed lines at the top panels show the ionized hydrogen density profiles.  
  In the fifth-row, the red solid and blue dashed lines show the charge of the large dust and the small dust, respectively.  
  In the bottom panels, the red solid and blue dashed lines show the relative velocity between the large dust and gas and between the small dust and gas, respectively.  
  Ionization front (IF) is indicated by vertical brown dashed lines and the shock front (SF) is indicated by vertical green dot-dot-dashed lines.
} 
\label{d2g} 
\end{figure*}

We present densities, gas temperature, dust-to-gas mass ratios for large and small grains, large-dust-to-small-dust mass ratios ($\rho_\mathrm{d, Large}/\rho_\mathrm{d, Small}$), the grain electric potential, and relative velocity between dust and gas as functions of radius in Fig~\ref{d2g}. 
In order to compare the simulation results on the dust size distribution with each other, we present the results at the time when the shock front reaches to $\sim$15~pc. 
In order to study the 
dependence of the dust size distribution on time and the luminosity of the radiation source, we also present the simulation result of Cloud~2 at $t=$1.1~My: the same irradiation time as Cloud~4. 
In the top panels, the hydrogen number density is indicated by the red solid lines and that of \ion{H}{ii} is indicated 
 by the blue dot-dashed lines. Initial states of the simulations are shown by black dotted lines. 
 In the fifth row, the charges of dust grains with size 0.1 \micron\, and 0.01 \micron\, are indicated by the red solid and blue dot-dashed lines, respectively.
 The black dotted lines show the initial profiles (i.e. 0~V). 
 In the bottom panels, the relative velocity between dust grains with size 0.1 \micron\, and gas and between dust grains with size 0.01 \micron\, and gas are indicated by the red solid and blue dot-dashed lines, respectively. 
Note that the radiation source becomes stronger from Cloud 2 to Cloud 4.

We find that radiation pressure affects the dust distribution within an \ion{H}{ii} region depending on the grain size. In Fig~\ref{d2g}, we divide them into the following four regions:  
\begin{enumerate}
 \item[(a)] From the central part, radiation pressure removes both large and small dust grains and creates a dust cavity (the yellow shaded region).
 \item[(b)] Within an \ion{H}{ii} region, $\rho_\mathrm{d, Large}/\rho_\mathrm{d, Small}$ has a peak. Between the region `a' and this peak, there is a region where $\rho_\mathrm{d, Large}/\rho_\mathrm{d, Small}$ takes the local minimum value (the cyan shaded region), for example, at $r\sim 4$~pc in Cloud~2 at $t=2.9$~Myr. 
 \item[(c)] The region that contains the peak mentioned above is shaded by magenta.
 \item[(d)] The $\rho_\mathrm{d, Large}/\rho_\mathrm{d, Small}$ is also reduced just behind the ionization front (the gray shaded region), 
 for example, at $r\sim 6$~pc in Cloud~2 at $t=2.9$~Myr. 
 \end{enumerate}

We find that the dust cavity radius becomes larger as the radiation source becomes brighter (the regions `a').  
The reasons are as follows. 
grain electric potential of the dust grains with the same size within $r=2$~pc is almost the same among all simulations and the number density of the gas becomes smaller for stronger radiation source. 
Since the dust drag force strongly depends on the grain electric potential, the number density of gas, and radiation pressure on 
dust, relative velocity between dust and gas becomes larger for the brighter source. 

In the region `b' and `d', the ratio $\rho_\mathrm{d, Large}/\rho_\mathrm{d, Small}$ is decreased from the initial condition
when the radiation source is a single OB star (Cloud~2). Except in the ionization front (vertical brown dashed lines)
which is contained in the region `d', radiation pressure preferentially removes large dust grains from these
regions.
The photoelectric yield of the large dust grains is smaller than that of the small dust grains, 
and hence grain electric potential of the large dust grains becomes smaller than that of the small dust grains. 
Coulomb drag between large dust grains
and gas therefore becomes weaker than that between small dust grains and gas.
On the other hand, since Cloud~4 has the strongest radiation source and hence it makes grain electric potentials largest among the simulations, 
the dust size segregation in the regions `b' and `d' is less prominent.
Even when we compare Cloud~2 and 4 at the same irradiation time, $t=$1.1~Myr,
 the dust size distributions inside \ion{H}{ii} regions are different. 
Luminosity of the radiation source must be the main cause of the dust size segregation. 

The ratio $\rho_\mathrm{d, Large}/\rho_\mathrm{d, Small}$ in all simulations has a peak in the regions `c'. Since dust grains have large negative charge in the regions `c' and `d', the coulomb drag force between dust and gas is strong and hence dust and gas are tightly coupled each other. 
Large dust grains are, therefore, removed from the regions `a' and `b' and gathered in the regions  `c'.

At the ionization front and the shock front (vertical green dot-dot-dashed lines), the relative velocity, $v_\mathrm{d} -
v_\mathrm{g}$ has downward peaks. In theses fronts, gas pressure force exceeds radiation pressure force. Since
the dust drag time depends on the dust grain size, the dust-gas relative velocity also depends on the grain
size. As a result, $\rho_\mathrm{d, Large}/\rho_\mathrm{d, Small}$ is slightly reduced in these fronts.

\section{Discussion and Conclusions}
\label{sec:Discussion}

We have investigated radiation feedback in dusty clouds by one-dimensional multi-fluid hydrodynamic simulations. 
In order to study 
spatial dust distribution inside \ion{H}{ii} regions, we solve gas and dust motion self-consistently. 
We also investigate dust size distribution within \ion{H}{ii} regions by 
considering dust grains with two different sizes. 

We find that radiation pressure creates dust cavity regions. 
We confirm that the size of the dust cavity region 
is broadly agree with the observational estimate \citep{Inoue2002}. 

We also find that radiation pressure preferentially removes large dust from \ion{H}{ii} regions 
in the case of a single OB star. 
This result is almost the same as in \cite{Akimkin2015}. 
The dust size distribution is, however, less affected when the radiation source is a star cluster, in other word, a more luminous case. 
Resulting dust size distributions largely depend on the luminosity of the radiation source. 


We assume dust is graphite. There are, however, other forms of dust such as silicate. 
Since the photoelectric yield and the absorption coefficient depend on a dust model,  
spatial dust distribution of dust grains may become different when we use a different dust model.
For example, since silicate has a larger work function and a smaller absorption coefficient 
than graphite, the cavity size in the silicate case may become larger than that in the graphite case (see \citealt{Akimkin2015, Akimkin2017} for details).


In our simulations, we neglect the effect of sputtering that changes the dust grain size. We estimate 
this according to \cite{Nozawa2006}, and confirm that sputtering effect is negligible in 
our simulations. However, if we consider the smaller dust grains, we may have to include 
the sputtering. 


\section*{Acknowledgements}

We are grateful to Takashi Kozasa, Takashi Hosokawa, and Shu-ichiro Inutsuka for helpful discussion. 
SI acknowledges Grant-in-Aid for JSPS Research Fellow (17J04872) 
and TO acknowledge the financial support of MEXT KAKENHI Grant (16H01085).
Numerical simulations were partly carried out with Cray XC30 in CfCA at NAOJ.




\bibliographystyle{mnras}
\bibliography{ishiki_2_clean.bbl}






\appendix

\section{Shock tube tests}
\label{sec:stt}

\begin{figure*}
\centering 
\includegraphics[width=13.5cm]{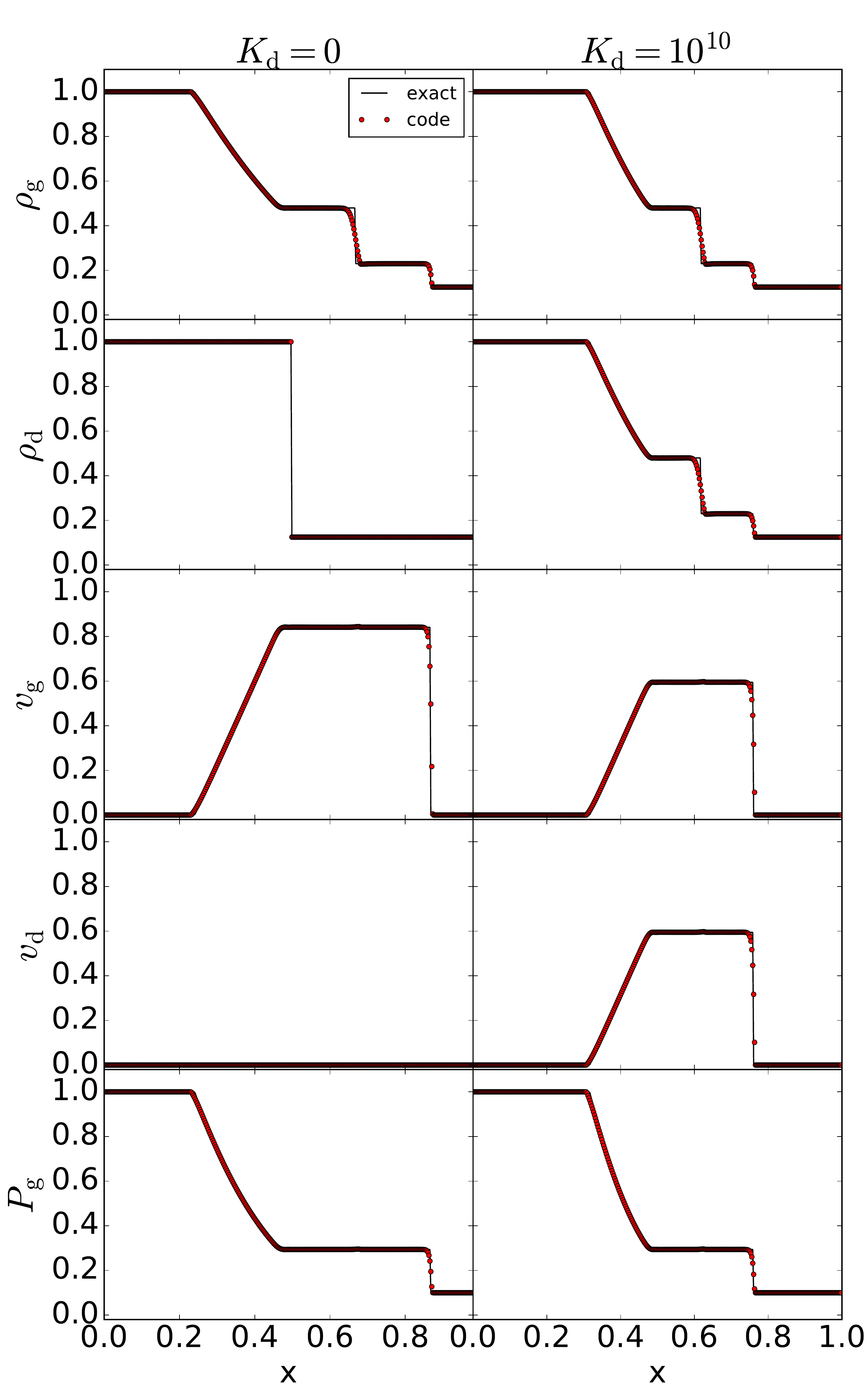} 
\caption{
  Gas mass density (top), dust mass density (second from the top), gas velocity (middle), 
  dust velocity (second from the bottom), and pressure (bottom) profiles.  
  In the left and right panels, we show the results for $K_\mathrm{d}=0$ and $K_\mathrm{d}=10^{10}$, respectively. 
  The red dots indicate the numerical results. 
  The black solid lines at the left panels represent the exact solutions for $K_\mathrm{d}=0$. 
  The black solid lines at the right panels represent the exact solutions for $K_\mathrm{d}=\infty$. 
  For both simulations, we show the results with the adiabatic index $\gamma$ = 1.67 at $t$ = 0.2.
} 
\label{test} 
\end{figure*}

In order to investigate whether our method is reliable, we perform shock tube tests. 
Since the effect of the dust becomes almost negligible in shock tube tests if we assume dust-to-gas mass ratio as 6.7$\times$10$^{-3}$ (the value we assume in our simulations) and hence we will not be able to investigate whether the numerical code is  reliable or not, we assume dust-to-gas mass ratio as 1 in the shock tube tests.
The initial condition of the shock tube problem is as follows:
\begin{eqnarray}
\rho_\mathrm{g} &= \rho_\mathrm{d} = 
\begin{cases}
1,\, &(x<0.5),\nonumber \\
0.125,\, &(x>0.5), \nonumber 
\end{cases} \\
P_\mathrm{g} &=
\begin{cases}
1,\, &(x<0.5),\nonumber \\
0.1,\, &(x>0.5), \nonumber 
\end{cases} \\
\gamma &= 1.67, & \nonumber
\end{eqnarray}
where $\gamma$ is heat capacity ratio. 
Since the analytic solutions are known for $K_\mathrm{d}=0$ and $\infty$, we perform test calculations for $K_\mathrm{d}=0$ and 
$K_\mathrm{d}=10^{10}$ ($\Delta t_\mathrm{sim} \gg (\rho_\mathrm{g} \rho_\mathrm{d})/(\rho_\mathrm{d}+\rho_\mathrm{g}) K_\mathrm{d}^{-1}\equiv t_\mathrm{d}$), 
where $\Delta t_\mathrm{sim}$ is the time scale of the shock tube problem and $t_\mathrm{d}$ is the drag stopping time. 
We use linearly spaced 400 meshes between $x=0$ and $1$. Time steps we use for these simulations are $\Delta t =2.5\times10^{-4}$ for $K_\mathrm{d}=0$ and $\Delta t=4.2\times10^{-4}$ for $K_\mathrm{d}=10^{10}$. 
The results are shown in Fig \ref{test}.   
We confirm that the numerical results agree with the analytic solutions. 

\section{dust grains with two sizes and gas dynamics}
\label{sec:2dust}

In order to investigate the spatial variation of the grain size distribution inside \ion{H}{ii} regions, we solve following hydrodynamics equations, where we consider dust grains with two sizes (dust-1 and dust-2):
\begin{eqnarray}
\frac{\partial}{\partial t} \rho_\mathrm{g} + \frac{\partial}{\partial x} \rho_\mathrm{g} v_\mathrm{g} &=& 0 \nonumber \\ 
\frac{\partial}{\partial t} \rho_\mathrm{d1} + \frac{\partial}{\partial x} \rho_\mathrm{d1} v_\mathrm{d1} &=& 0 \nonumber \\ 
\frac{\partial}{\partial t} \rho_\mathrm{d2} + \frac{\partial}{\partial x} \rho_\mathrm{d2} v_\mathrm{d2} &=& 0 \nonumber \\ 
\frac{\partial}{\partial t} \rho_\mathrm{g} v_\mathrm{g} + \frac{\partial}{\partial x} \rho_\mathrm{g} v_\mathrm{g}^2 &=& \rho_\mathrm{g} a_\mathrm{gra} + f_\mathrm{rad,g} - \frac{\partial}{\partial x} P_\mathrm{g} \nonumber \\ 
& & + K_\mathrm{d1} (v_\mathrm{d1} - v_\mathrm{g})  + K_\mathrm{d2} (v_\mathrm{d1} - v_\mathrm{g}) \nonumber \\
\frac{\partial}{\partial t} \rho_\mathrm{d1} v_\mathrm{d1} + \frac{\partial}{\partial x} \rho_\mathrm{d1} v_\mathrm{d1}^2 &=&  \rho_\mathrm{d1} a_\mathrm{gra} + f_\mathrm{rad,d1}  \nonumber \\ 
& &  + K_\mathrm{d1} (v_\mathrm{g} - v_\mathrm{d1}) \nonumber \\ 
\frac{\partial}{\partial t} \rho_\mathrm{d2} v_\mathrm{d2} + \frac{\partial}{\partial x} \rho_\mathrm{d2} v_\mathrm{d2}^2 &=&  \rho_\mathrm{d2} a_\mathrm{gra} + f_\mathrm{rad,d2}  \nonumber \\ 
& & + K_\mathrm{d2} (v_\mathrm{g} - v_\mathrm{d2}) \nonumber \\ 
\frac{\partial}{\partial t} \left( \frac{1}{2} \rho_\mathrm{g} v_\mathrm{g}^2 + e_\mathrm{g} \right) &+& \frac{\partial}{\partial t} \left( \frac{1}{2} \rho_\mathrm{d1} v_\mathrm{d1}^2 +  \frac{1}{2} \rho_\mathrm{d2} v_\mathrm{d2}^2  \right) \nonumber \\
+ \frac{\partial}{\partial x} \left( \frac{1}{2} \rho_\mathrm{g} v_\mathrm{g}^2 + h_\mathrm{g} \right) v_\mathrm{g} &+& \frac{\partial}{\partial x} \left(  \frac{1}{2} \rho_\mathrm{d1} v_\mathrm{d1}^3 + \frac{1}{2} \rho_\mathrm{d2} v_\mathrm{d2}^3 \right)  \nonumber \\
&=& \left( \rho_\mathrm{g} v_\mathrm{g} +  \rho_\mathrm{d1} v_\mathrm{d1} +  \rho_\mathrm{d2} v_\mathrm{d2}  \right) a_\mathrm{gra} \nonumber \\
& &  + f_\mathrm{rad,g} v_\mathrm{g} + f_\mathrm{rad,d1} v_\mathrm{d1}+ f_\mathrm{rad,d2} v_\mathrm{d2} \nonumber 
\end{eqnarray}
where $\rho_\mathrm{d1}$ is the mass density of dust-1, $\rho_\mathrm{d2}$ is the mass density of dust-2, $v_\mathrm{d1}$ is 
the velocity of dust-1, $v_\mathrm{d2}$ is the velocity of dust-2, $f_\mathrm{rad,d1}$ is the radiation pressure gradient force on dust-1, $f_\mathrm{rad,d2}$ is the radiation pressure gradient force on dust-2, $K_\mathrm{d1}$ is the drag coefficient between gas and dust-1, and $K_\mathrm{d2}$ is the drag coefficient between gas and dust-2. 

In order to solve dust drag force stably, we use following algorithm for the equation of momentum: 
\begin{equation}
\begin{split}
\begin{bmatrix}
p_\mathrm{d1}^{*} \left( = \rho_\mathrm{d1}^\mathrm{t+\Delta t} v_\mathrm{d1}^*  \right) \\
p_g^{*}\left( = \rho_\mathrm{g}^\mathrm{t+\Delta t} v_\mathrm{g}^*  \right) \\
p_\mathrm{d2}^{*} \left( = \rho_\mathrm{d2}^\mathrm{t+\Delta t} v_\mathrm{d2}^*  \right) \\
\end{bmatrix}
&=
\begin{bmatrix}
p_\mathrm{d1}^t \\
p_g^t \\
p_\mathrm{d2}^t
\end{bmatrix}
+ \begin{bmatrix}
F_\mathrm{p, d1}(\rho_\mathrm{d1}^t,v_\mathrm{d1}^t)  \\
F_\mathrm{p, g}(\rho_g^t,v_\mathrm{g}^t,e_g^t) \\
F_\mathrm{p, d2}(\rho_\mathrm{d2}^t,v_\mathrm{d2}^t)  \\
\end{bmatrix}
\Delta t,\label{cdv3}
\end{split}
\end{equation}
\begin{equation}
\begin{split}
& 
\begin{bmatrix}
p_\mathrm{d1}^\mathrm{t+\Delta t} \\
p_g^\mathrm{t+\Delta t} \\
p_\mathrm{d2}^\mathrm{t + \Delta t}
\end{bmatrix}  \\
&=
\begin{bmatrix}
\rho_\mathrm{d1}^{t+\Delta t} \\
\rho_\mathrm{g}^{t+\Delta t} \\
\rho_\mathrm{d2}^{t+\Delta t}
\end{bmatrix} 
\frac{ p_\mathrm{d1}^{*} + p_\mathrm{g}^{*} + p_\mathrm{d2}^{*} }{ \rho_\mathrm{d1}^{t+\Delta t} + \rho_g^{t+\Delta t} + \rho_\mathrm{d2}^{t+\Delta t} }  \\
&+
\begin{bmatrix}
\rho_\mathrm{d1}^\mathrm{t+\Delta t} \\
\rho_g^\mathrm{t+\Delta t} \\
\rho_\mathrm{d2}^\mathrm{t+\Delta t}
\end{bmatrix}
\left[ a_\mathrm{gra} + \frac{f_\mathrm{d1} + f_g + f_\mathrm{d2}}{\rho_\mathrm{d1}^\mathrm{t+\Delta t} + \rho_g^\mathrm{t+\Delta t} + \rho_\mathrm{d2}^\mathrm{t+\Delta t} }\right] \Delta t \\
&+ 
\begin{bmatrix}
\frac{b}{a+x} \\
1 \\
\frac{c}{d+x}
\end{bmatrix}
\frac{\rho_g \mathrm{e}^{x\Delta t}}{x(x-y)} \left[ b(d+x) v_\mathrm{d1}^* + (a+x)(d+x)v_g^* + c(a+x)v_\mathrm{d2}^* \right]  \\
&+ 
\begin{bmatrix}
\frac{b}{a+y} \\
1 \\
\frac{c}{d+y}
\end{bmatrix}
\frac{\rho_g \mathrm{e}^{y\Delta t}}{y(y-x)} \left[ b(d+y) v_\mathrm{d1}^* + (a+y)(d+y)v_g^* + c(a+y)v_\mathrm{d2}^* \right]  \\
&+ 
\begin{bmatrix}
\frac{b}{a+x} \\
1 \\
\frac{c}{d+x}
\end{bmatrix}
\frac{\mathrm{e}^{x\Delta t}-1}{x^2(x-y)} \left[ a(d+x) f_\mathrm{d1} + (a+x)(d+x)f_g + d(a+x)f_\mathrm{d2} \right]  \\
&+ 
\begin{bmatrix}
\frac{b}{a+y} \\
1 \\
\frac{c}{d+y}
\end{bmatrix}
\frac{\mathrm{e}^{y\Delta t}-1}{y^2(y-x)} \left[ a(d+y) f_\mathrm{d1} + (a+y)(d+y)f_g + d(a+y)f_\mathrm{d2} \right], 
\label{cdv4}
\end{split}
\end{equation}
where $f_\mathrm{d1}$ is  the force on dust-1 ($f_\mathrm{d1}=f_\mathrm{rad,d1}$), $f_\mathrm{d2}$ is  the force on dust-2 ($f_\mathrm{d2}=f_\mathrm{rad,d2}$), 
\begin{equation}
\begin{split}
a &= \frac{K_\mathrm{d1}}{\rho_\mathrm{d1}^\mathrm{t+\Delta t}}, \nonumber \\
b &= \frac{K_\mathrm{d1}}{\rho_\mathrm{g}^\mathrm{t+\Delta t}}, \nonumber \\
c &= \frac{K_\mathrm{d2}}{\rho_\mathrm{g}^\mathrm{t+\Delta t}}, \nonumber \\
d &= \frac{K_\mathrm{d2}}{\rho_\mathrm{d2}^\mathrm{t+\Delta t}}, \nonumber \\
x &= -\frac{1}{2} \left[ (a+b+c+d) + \sqrt{(a+b+c+d)^2 - 4(ad+ac+bd)} \right], \nonumber \\
\end{split}
\end{equation}
and
\begin{equation}
y = \frac{  (ad+ac+bd) }{ x }. \nonumber 
\end{equation}
As in section \ref{subsubsec:d1}, in order to determine the relative velocity between gas and dust, we use equation (\ref{cdv4}) which is the 
exact solution of the following equations:
\begin{equation}
\begin{split}
\rho_\mathrm{d2} \frac{d}{dt} v_\mathrm{d2} &= f_\mathrm{rad,d2}  + \rho_\mathrm{d2} a_\mathrm{gra} + K_\mathrm{d2} (v_\mathrm{g} - v_\mathrm{d2}), \\
\rho_\mathrm{g} \frac{d}{dt} v_\mathrm{g} &= f_\mathrm{rad,g} - \frac{\partial}{\partial x}P_\mathrm{g} + \rho_\mathrm{g} a_\mathrm{gra}  \\
&+ K_\mathrm{d1} (v_\mathrm{d1} - v_\mathrm{g})+ K_\mathrm{d2} (v_\mathrm{d2} - v_\mathrm{g}), \label{dve2} \\
\rho_\mathrm{d1} \frac{d}{dt} v_\mathrm{d1} &= f_\mathrm{rad,d1}  + \rho_\mathrm{d1} a_\mathrm{gra} + K_\mathrm{d1} (v_\mathrm{g} - v_\mathrm{d1}). 
\end{split}
\end{equation}
In order to solve momentum equations, we therefore first solve the momentum advection (\ref{cdv3}), and then we solve the exact solution of the equation (\ref{dve2}) by equation (\ref{cdv4}).

In the case for $|x| \Delta t \ll 1$ or $|y| \Delta t \ll 1$, we use Taylar expansion, $\mathrm{e}^{x\Delta t} \approx 1 + x\Delta t$ or $\mathrm{e}^{y\Delta t} \approx 1 + y\Delta t$, and prevent the numerical error in calculating $(\mathrm{e}^{x\Delta t}-1)/x$ from becoming too large. 

\section{The terminal velocity approximation}

We here show that the terminal velocity approximation may give an unphysical result when 
the simulation time step $\Delta t$ is shorter  than the drag stopping time $t_\mathrm{d}$.
In order to derive dust velocity and gas velocity, we have used the equation~(\ref{cdv2}). 
On the other hand, \cite{Akimkin2017} used the terminal velocity approximation. 
When we employ the terminal velocity approximation, the equation~(\ref{cdv2}) 
is transforms into the following form: 
\begin{equation}
\begin{split}
\begin{bmatrix}
p_\mathrm{d}^\mathrm{t+\Delta t} \\
p_\mathrm{g}^\mathrm{t+\Delta t}
\end{bmatrix}
&=
\begin{bmatrix}
\rho_\mathrm{d}^\mathrm{t+\Delta t} \\
\rho_\mathrm{g}^\mathrm{t+\Delta t} 
\end{bmatrix}
\frac{ p_\mathrm{d}^*+p_\mathrm{g}^* }{ \rho_\mathrm{d}^\mathrm{t+\Delta t} + \rho_g^\mathrm{t+\Delta t}  } 
+
\begin{bmatrix}
\rho_\mathrm{d}^\mathrm{t+\Delta t} \\
\rho_\mathrm{g}^\mathrm{t+\Delta t} 
\end{bmatrix}
 a_\mathrm{gra} \Delta t  \\
&+ 
\begin{bmatrix}
(\rho_\mathrm{d}^\mathrm{t+\Delta t} \Delta t + \rho_\mathrm{g}^\mathrm{t+\Delta t} t_\mathrm{d})f_\mathrm{d} \\
(\rho_\mathrm{g}^\mathrm{t+\Delta t} \Delta t + \rho_\mathrm{d}^\mathrm{t+\Delta t} t_\mathrm{d})f_\mathrm{g}
\end{bmatrix} 
\frac{1}{\rho_\mathrm{g}^\mathrm{t+\Delta t} + \rho_\mathrm{d}^\mathrm{t+\Delta t}}
\\
& +
\begin{bmatrix}
\rho_\mathrm{d}^\mathrm{t+\Delta t} f_\mathrm{g} \\
\rho_\mathrm{g}^\mathrm{t+\Delta t} f_\mathrm{d}
\end{bmatrix}  
 \frac{(\Delta t - t_\mathrm{d})}{\rho_\mathrm{g}^\mathrm{t+\Delta t} + \rho_\mathrm{d}^\mathrm{t+\Delta t}} .  
\end{split} \label{cdv2ap}
\end{equation}
The advantage of the equation (\ref{cdv2}) is that it is accurate even for 
$\Delta t < t_\mathrm{d}$. In contrast, the equation~(\ref{cdv2ap}) becomes 
inaccurate for $\Delta t \ll t_\mathrm{d}$, since the relation 
of $\Delta t$ and $t_\mathrm{d}$ 
should be $\Delta t \gg t_\mathrm{d}$ 
in order the terminal velocity approximation to be valid. 
For example, the direction of $f_\mathrm{d}$ on gas and 
that of $f_\mathrm{g}$ on dust in the equation~(\ref{cdv2ap}) 
becomes opposit for $\Delta t < t_\mathrm{d}$.

We perform simulations by using equation~(\ref{cdv2ap}) in stead of equation~(\ref{cdv2}) and 
compare the simulation results. 
Simulation results do not largely change for Cloud~2, 3, and 4. 
The numerical simulation of Cloud~1, however, is crashed, 
since the timestep becomes $\Delta t \ll t_\mathrm{d}$ at some
steps.

\begin{table}
  \centering
  \caption{
  Numerical setup for the IGM, the \ion{H}{ii} region, and the \ion{H}{i} region. 
    The number densities of hydrogen and ionized hydrogen are represented by ${n}_\mathrm{H}$ 
    and $n_\mathrm{\ion{H}{ii}}$. 
    The temperature of gas is represented by $T_\mathrm{g}$. 
    The radius of a grain is represented by $a_\mathrm{dust}$. 
    The grain electric potential of dust grains is represented by $V_\mathrm{d}$.  
    The relative velocity between a dust grain and gas is represented by $\Delta v$.
  }
\begin{tabular}{lcccccc} \hline
  \, & ${n}_\mathrm{H}$ & ${n}_\mathrm{\ion{H}{ii}}$ & ${T}_\mathrm{g}$ & ${a}_\mathrm{dust}$ & ${V}_\mathrm{d}$ & $\Delta v$\\ 
  \, & (cm$^{-3}$) & (cm$^{-3}$) & (K) &  (\micron)  & (V) & (km s$^{-1}$) \\ \hline
  IGM & 10$^{-5}$ & 10$^{-5}$ & 10$^4$ & 0.1 & 20 & 0 \\ 
  \ion{H}{ii} region & 10 & 10 & 10$^4$ & 0.1 & 5 & 0 \\ 
  \ion{H}{i} region &10$^2$  & 0 & 10$^2$ & 0.1 & 0 & 0 \\ \hline
   \end{tabular}
  \label{TabC2}
\end{table}

\begin{figure}
\centering 
\includegraphics[width=8cm]{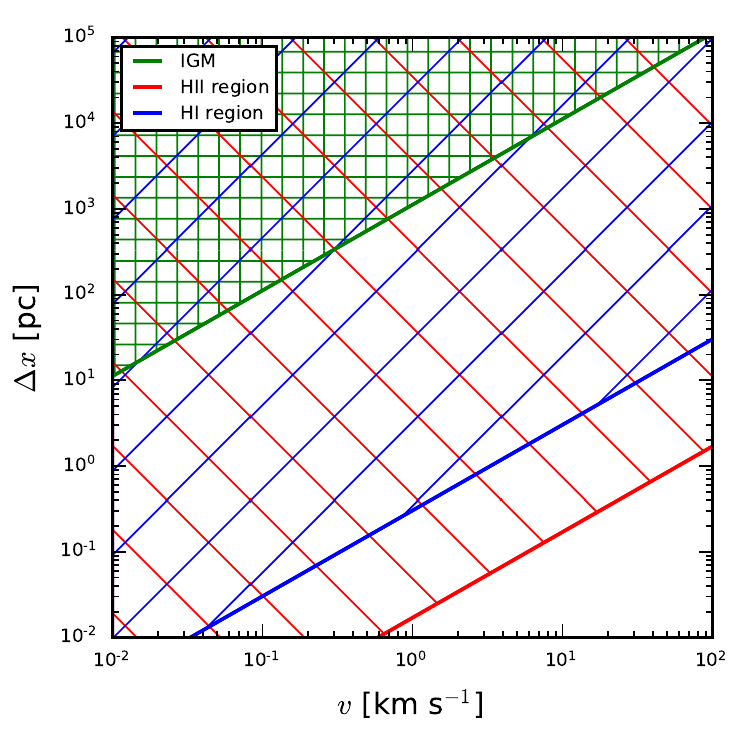} 
\caption{
The green, red, and blue hatched regions represent the condition of $t_\mathrm{CFL} > t_\mathrm{d}$ for the IGM, 
the \ion{H}{ii} region, and the \ion{H}{i} region, respectively. 
} 
\label{timestep_ch} 
\end{figure}

The relation between $\Delta t$ and $t_\mathrm{d}$ becomes $\Delta t < t_\mathrm{d}$ when the drag stopping time $t_\mathrm{d}$ is larger than the chemical timestep $\Delta t_\mathrm{chem}$ or the timestep $\Delta t_\mathrm{CFL}$ defined by Clourant-Friedrichs-Lewy condition. The chemical timestep is defined in equation (7) in paper I. 

In Fig \ref{timestep_ch}, we present the condition for $t_\mathrm{CFL}(\equiv \alpha \Delta x/v) > t_\mathrm{d}$ in the case of the intergalactic medium (IGM), the \ion{H}{ii} region, and the \ion{H}{i} region, where $\alpha$ is constant (we assume $\alpha=0.1$), $\Delta x$ is the mesh size, and $v$ is velocity. The details of numerical setup for the IGM, the \ion{H}{ii} region, and the \ion{H}{i} region are listed in Tab \ref{TabC2}.  
The green, red, and blue hatched regions represent the condition for $t_\mathrm{CFL} > t_\mathrm{d}$ for the IGM, the \ion{H}{ii} region, and the \ion{H}{i} region, respectively. 
If the relation between $t_\mathrm{CFL}$ and $t_\mathrm{d}$ becomes $t_\mathrm{CFL} \ll t_\mathrm{d}$, the simulation may become unstable.

\bsp	
\label{lasPHage}
\end{document}